\documentclass[12pt]{article}
\begin{document}
{\large\bf Bianchi type IX world with torsion and shear}

\vspace{2cm}

\noindent
\noindent
\begin{center}
{\large L. C. Garcia de Andrade\footnote{garcia@dft.if.uerj.br -\
Departamento de Fisica Teorica-Instituto de F\'{\i}sica -UERJ, Rua S\~{a}o
Francisco Xavier 524, Maracan\~{a},R.J
Cep: 20550 - Rio de Janeiro, Brasil.} and G. A.
Monerat\footnote{monerat@if.uff.br -\ Universidade Federal Fluminense,
Instituto de F\'{\i}sica,
Av., Litr\^{o}nea s/n - Boa Viagem, Cep:24210-340,  Niter\'{o}i ,R.J,
Brasil.}}
\end{center}
\vspace{1cm}
\begin{center}
\large Abstract
\end{center}
Oscillating phases in Bianchi type IX anisotropic universes are obtained as
linearized solutions of Einstein-Cartan theory of gravity.Instead of making
use of Weyssenhoff spinning fluids we addopt axionic torsion matter where
the torsion tensor is totally skew-symmetric as in Kalb-Rammond type
theories.Spin-torsion,metric and matter density oscillate at different
phases.
An upper bound oscillatory limit for torsion is obtained from the model.
Metric
coefficients correspond to two harmonic oscillators equations,the first
describes a harmonic simple oscillator and the second a forced harmonic
oscillator.
\vspace{1cm}

Recently Harko et al \cite{1} and H.Q.Lu and K.S.Cheng \cite{2} have
investigated in detail the Bianchi types I and V cosmological models in the
realm of Einstein-Cartan theory of gravity \cite{3}. In this letter a
linearized solution of Einstein-Cartan (EC) theory representing a Bianchi
type IX \cite{4} cosmological model is presented.The solution possess the
interesting feature of being an oscillating solution where the spin-torsion
and matter densities oscillate at distinct phasess as well as the metric
coefficients. Oscillating universes are important in the reheating phases of
the Universe just after inflation. In fact this work in a certain sense is a
continuation of our previous paper \cite{4} on inflationary phases as a de
Sitter solution of EC gravity.Anisotropic solutions could also lead to
topological defects in EC gravity.The physical motivation of using
anisotropic models is according to Harko et al \cite{1} be due to the fact
that the spin is polarized. Since we only deal with linearized solution the
structure of singularities on these models is out the scope of this
work.Nevertheless is important to mention that the vacuum Bianchi type IX
exact solution of Einstein field equation of General Relativity is already
non-singular. Let us now consider the geometry of Bianchi type IX
cosmological model is given by \cite{5}
\begin{equation}
ds^{2}=2{\omega}^{1}dt+g_{11}({\omega})^{2}+e^{2{\beta}}( {\omega}^{2}+
{\omega}^{3})
\label{1}
\end{equation}

\noindent
where Taub space corresponds to $g_{11}>0$ and the NUT space corresponds to
$g_{11}<0$.The coordinate $t$ can be parametrized so that $g_{01}=1$ and in
the basis
\begin{equation}
{\sigma}_{1}={\omega}^{1}
\label{2}
\end{equation}
and ${\sigma}^{a}=e^{{\beta}}{\omega}^{a}$ where ${a=1,2}$ and
$d{\sigma}^{2}=
{a}_{S}{{\omega}^{S}}^{2}$ is the hypersurface metric.Making $g=g_{11}$ the
we are left with a off-diagonal metric.Making use of an appropriate frame of
differential forms the following components of the Riemannian Einstein
tensor is
\begin{equation}
G_{00}=R_{00}
\label{3}
\end{equation}
and
\begin{equation}
G_{11}=-\frac{1}{2}R
\label{4}
\end{equation}
where $R$ is the Ricci scalar and $G_{22}=G_{33}$.Component $G_{01}$ is also
distinct from zero.Taking the axionic torsion energy-stress tensor as
\begin{equation}
{{T}^{Ax}}^{l}_{i}=3S_{ijk}S^{ljk}-\frac{1}{2}{\delta}^{l}_{i}S^{2}
\label{5}
\end{equation}
where $S^{2}=S_{ijk}S^{ijk}$ is the the spin-torsion polarized density from
the non-symmetric connection and ${i,j,k=0,1,2,3}$.Here we assume that the
only nonvanishing component of the torsion tensor is$S_{012}$.The matter
stress-energy tensor is given by
\begin{equation}
T_{ij}={\rho}u_{i}u_{j}
\label{6}
\end{equation}
which corresponds to the dust pressureless axionic fluid where the four
velocity is given by
\begin{equation}
u^{i}={\delta}^{i}_{0}+{\delta}^{i}_{1}
\label{7}
\end{equation}
With these tools in hands one may compute the Einstein-Cartan equations as
\begin{equation}
C(t)+\frac{1}{2}e^{-4{\beta}}={\rho}+S^{2}
\label{8}
\end{equation}
and
\begin{equation}
g^{2}C(t)-\frac{1}{4}g{\ddot{g}}-\frac{1}{2}ge^{-2{\beta}}+\frac{1}{4}g^{2}e
^{-4}-\frac{1}{4}{\dot{g}}{\dot{\beta}}={\rho}(1+g)+S^{2}
\label{9}
\end{equation}
where $C(t)=-2({\ddot{\beta}}+{\dot{\beta}}^{2})$.The remaining EC equations
are
\begin{equation}
-g{\ddot{\beta}}+\frac{1}{2}e^{-2{\beta}}-\frac{1}{4}e^{-4{\beta}}-\frac{1}{
4}{\ddot{g}}=(1-\frac{g}{2})S^{2}
\label{10}
\end{equation}
and
\begin{equation}
-g{\ddot{\beta}}+\frac{3}{4}g e^{-4{\beta}}+gC-\frac{1}{2}{e^{-2{\beta}}}=
{\rho}(1+g)-\frac{{S}^{2}}{2}
\label{11}
\end{equation}

\noindent
These equations are clearly very complicated to be solved without
computers.Since for our physical purposes it is enough to deal with the
linear approximation of the EC equations one may reduce the above EC
equations to
\begin{equation}
-2{\ddot{\beta}}+\frac{1}{2}e^{-4{\beta}}={\rho}+S^{2}
\label{12}
\end{equation}
and
\begin{equation}
-\frac{1}{2}g={\rho}(1+g)+\frac{1}{2}S^{2}
\label{13}
\end{equation}
and the remaining two equations are
\begin{equation}
\frac{1}{2}e^{-2{\beta}}-\frac{1}{4}g+\frac{1}{4}{\ddot{g}}=S^{2}
\label{14}
\end{equation}
and
\begin{equation}
-\frac{3}{4}g={\rho}(1+g)-\frac{1}{2}S^{2}
\label{15}
\end{equation}
After some algebra we reduce these last four field equations to
\begin{equation}
{\ddot{\beta}}+{\beta}= - \frac{5}{8}g
\label{16}
\end{equation}
and
\begin{equation}
{\ddot{g}}+4g=0
\label{17}
\end{equation}
The last equation is a very simple differential equation describing an
harmonic oscillator with constant frequency which yields the solution
\begin{equation}
g=cos2t
\label{18}
\end{equation}
which gives the first hint of the oscillatory behaviour of this
space-time.If the space will be of Taub type $g=g_{11}>0$ or of NUT type
$g=g_{11}<0$ will depend upon the time argument.Substitution of this
solution on the remaining equations of the system yields
\begin{equation}
{\rho}=\frac{\frac{1}{8}cos2t}{1+cos2t}
\label{19}
\end{equation}
we end up also with a slightly more complicated equation for the
${\beta}(t)$
metric function such as
\begin{equation}
{\ddot{\beta}}+{{\omega}_{0}}^{2}{\beta}=f(t)
\label{20}
\end{equation}
this represents an equation for the forced harmonic oscillator where the
square of the frequency is
\begin{equation}
{{\omega}_{0}}^{2}=\frac{3}{2}
\label{21}
\end{equation}
and the force $f(t)$ reads
\begin{equation}
f(t)=\frac{5}{8}cos2t
\label{22}
\end{equation}
The equation (\ref{20}) solves to
\begin{equation}
{\beta}=\frac{1}{4}cos2t
\label{23}
\end{equation}
.From expression (\ref{19}) one obtains immeadiatly a upper bound limit for
torsion $S^{2}<{\frac{1}{8}cos2t}$ by simply imposing the positivity of the
matter density ${\rho}>0$.Here we have consider approximation by dropping
terms like $g^{2}$.This shows that the spin-torsion density is also
oscillatory.Substitution of (\ref{23}) into equation (\ref{16}) allow us to
determine $S^{2}=\frac{1+\frac{19}{8}cos2t}{1+cos2t}$. An exact solution of
this model can be obtained in the near future along with a detailed
investigation of the singularity problem in Riemann-Cartan space for this
Bianchi Type IX model.This in fact is not the first time that oscillations
in the Universe appears in non-Riemannian cosmology.In 1983 Nurgaliev and
Ponomariev \cite{6} have investigated the evolution of the early universe in
EC gravity where the problems of gravitational instabilities have been
considered, wherethe solution seems to be stable to small homogeneous
perturbations and an oscillatory phase is found before inflation.Two
graphics are displayed on this text.The first two represent the behaviour of
the metric factors g and ${\beta}$ which allows us to say that the it
behaves oscillatory while its amplitude is distinct in different directions.
This behaviour is similar to the one obtained by Novikov and Zeldovich
\cite{6} who investigated the evolutionary  stages of the Bianchi type IX
universe in spacetimes without torsion. The behaviour for the square of
torsion is obtained in graphic $3$. More recently we show that teleparallel
spaces can induce a cosmological model where oscillation on the cosmic scale
factor are present. In this case the frequency of oscillations are damped by
the presence of torsion. In these two cases contrary to what happens here
both metrics are isotropic. One another important feature of the present
work is that our metric presents shear besides rotation. A more  detailed
investigation of oscillatory cosmological models in Riemann-Cartan
space-times and its implications to inflation may appear elsewhere.
\begin{flushleft}
{\Large Acknowledgements}
\end{flushleft}
\vspace{1cm}
 We would like to express our gratitude to $CNP_{q}$.(Brazil)
from financial support. I also thank Universidade do Estado do Rio de
Janeiro for partial financial support. Thanks are also due to Prof.H.P.de
Oliveira for helpful discussions on the subject of this paper and to Prof.
Yuri N. Obukhov for point it out an error in the first draft of this paper.

\vspace{5.0cm}

\noindent
\begin{figure}[h!]
\begin{minipage}[htb]{0.30\textwidth}
\special{eps:fig101.eps x=1.7 in y=1.5 in}
\caption{metric factor-g}
\label{f1}
\end{minipage}
\hspace{0.3cm}
\begin{minipage}[h!]{0.30\textwidth}
\special{eps:fig201.eps x=1.7 in y=1.5 in}
\caption{metric factor-${\beta}$}
\label{f2}
\end{minipage}
\hspace{0.3cm}
\begin{minipage}[htb]{0.30\textwidth}
\special{eps:fig301.eps x=1.7 in y=1.5 in}
\caption{Torsion decay}
\label{f3}
\end{minipage}
\end{figure}

\newpage
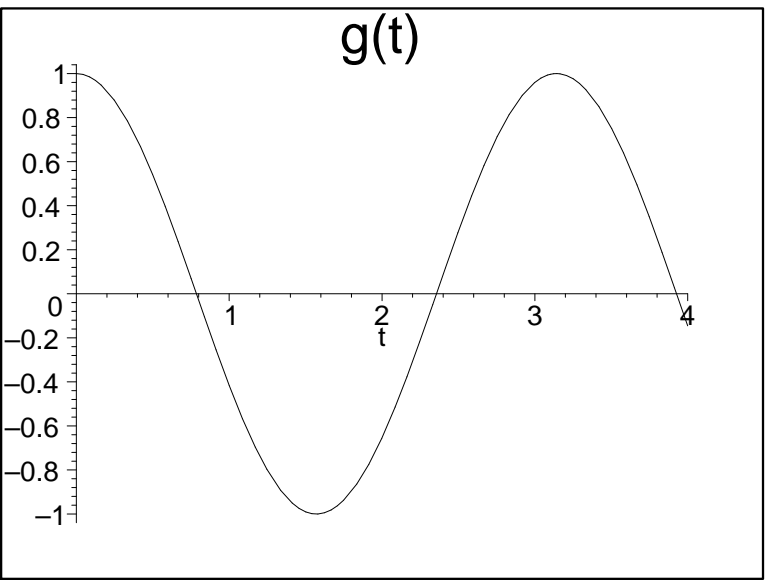
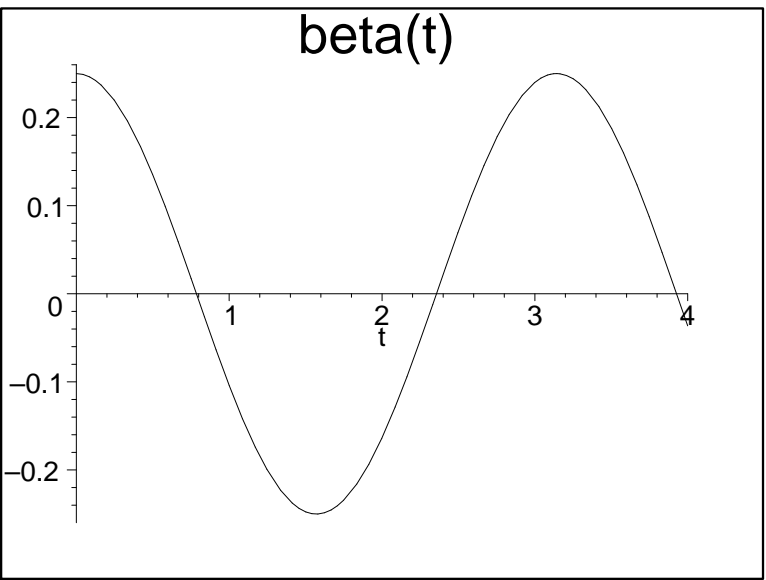
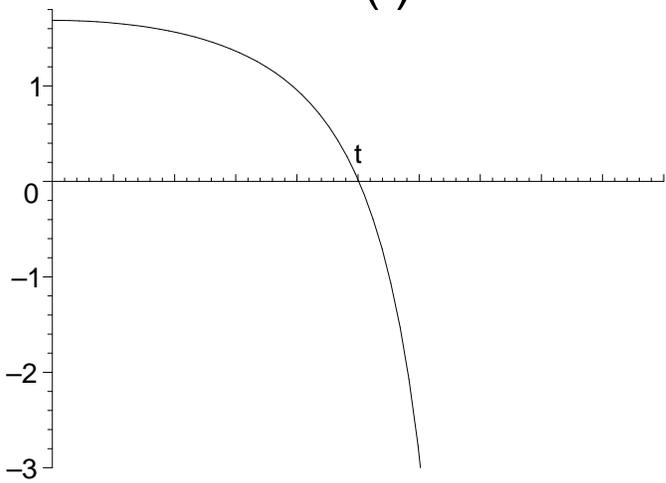
\end{document}